\documentclass{elsart}

\usepackage{graphicx}
\usepackage{amssymb}

\begin{document}

\begin{frontmatter}

\title{Trading activity as driven Poisson process: comparison with
empirical data}

\author{V. Gontis\corauthref{cor1}},
\ead{gontis@itpa.lt}
\ead[url]{http://www.itpa.lt/~gontis}
\author{B. Kaulakys},
\author{J. Ruseckas}
\corauth[cor1]{Corresponding author.}
\address{Institute of Theoretical Physics and Astronomy of Vilnius
University, A.~Go\v{s}tauto 12, LT-01108 Vilnius, Lithuania }

\begin{abstract}
We propose the point process model as the Poissonian-like stochastic sequence
with slowly diffusing mean rate and adjust the parameters of the model to the
empirical data of trading activity for 26 stocks traded on NYSE. The proposed
scaled stochastic differential equation provides the universal description of
the trading activities with the same parameters applicable for all stocks.
\end{abstract}

\end{frontmatter}

\section{Introduction}

A possible way to avoid the inconsistence of geometric Brownian
motion as a model of speculative markets is the assumption that
the volatility is itself a time-depending random variable. Within
this assumption there exist the discrete ARCH and GARCH models
\cite{Engle,Bollersev,Baillie,Engle2} as well as the continuous
stochastic volatility models \cite{Perello,Perello2,Yakovenko}. It
is empirically established that the autocorrelation function of
volatility decays very slowly with time exhibiting two different
exponents of the power-law distribution: for short time scales and
for the long time ones. However, common stochastic volatility
models are able reproduce only one time scale exponential decay
\cite{Perello2,White}. There is empirical evidence that the
trading activity is a stochastic variable with the power-law
probability distribution function (PDF) \cite{Mandelbrot, Lux} and
the long-range correlations \cite{Engle2, Plerou, Gabaix}
resembling the power-law statistical properties of volatility.
Empirical analysis confirms that the long-range correlations in
volatility arise due to those of the trading activity
\cite{Plerou}. On the other hand, the trading activity can be
modeled as the event flow of the stochastic point process with
more evident microscopic interpretation of the observed power-law
statistics.

Recently, we proposed the stochastic model of the trading activity in the
financial markets as Poissonian-like process driven by the stochastic
differential equation (SDE) \cite{ref13,PhysA2007}. Here we present the
detailed comparison of the model with the empirical data of the trading
activity for 26 stocks traded on NYSE. This enables us to present a more
precise model definition based on the scaled equation, universal for all
stocks. The proposed form of the difference equation for the intertrade time
can be interpreted as a discrete iterative description of the proposed model,
based on SDE.

\section{Stochastic model of trading activity}

We consider trades in the financial market as identical point events.  Such
point process is stochastic and defined by the stochastic interevent time
$\tau_{k}=t_{k+1}-t_{k}$, with $t_{k}$ being the occurrence times of the
events.  Recently we proposed \cite{ref13,PhysA2007} to model the flow of
trades in the financial markets as Poissonian-like process driven by the
multiplicative stochastic equation, i.e.\ we define the rate $n=1/\tau$ of this
process by the continuous stochastic differential equation
\begin{equation}
\mathrm{d}
n=\sigma^2\left[(1-\gamma_{\sigma})+\frac{m}{2}
\left(\frac{n_{0}}{n}\right)^{m}\right]\frac{n^4}{(n\epsilon+1)^2}\mathrm{d}
t+\frac{\sigma n^{5/2}}{(n\epsilon+1)}\mathrm{d}W.
\label{eq:nstoch2}
\end{equation}
This SDE with the Wiener noise $W$ describes the diffusion of the
stochastic rate $n$ restricted in some area: from the side of the
low values  by the term $m(n_{0}/n)^{m}/2$ and from the side of
high values by the relaxation $\gamma_{\sigma}$. The general
relaxation factor $\sigma^2n^4/(n\epsilon+1)^2\mathrm{d}t$ is
keyed with multiplicative noise $\sigma
n^{5/2}/(n\epsilon+1)\mathrm{d}W$ to ensure the power-law
distribution of $n$. The multiplicative noise is combined of two
powers to ensure the spectral density of $n$ with two power law
exponents. This form of the multiplicative noise helps us to model
the empirical probability distribution of the trading activity, as
well. A parameter $\epsilon$ defines the crossover between two
areas of $n$ diffusion. For more details see
\cite{ref13,PhysA2007}. Equation (\ref{eq:nstoch2}) has to model
stochastic rate $n$ with two power-law statistics, i.e., PDF and
power spectral density or autocorrelation, resembling the
empirical data of the trading activity in the financial markets.

We will analyze the statistical properties of the trading activity
$N(t,\tau_{\mathrm{d}})$ defined as integral of $n$ in the selected time window
$\tau_{\mathrm{d}}$,
\begin{equation}
N(t,\tau_{\mathrm{d}})=\int_{t}^{t+\tau_{\mathrm{d}}}n(t^{\prime})\mathrm{d}
t^{\prime} .
\end{equation}
The Poissonian-like sequence of trades described by the intertrade times
$\tau_{k}$ can be generated by the conditional probability
\begin{equation}
\varphi(\tau_{k}|n)=n\exp(-n\tau_{k}).
\label{eq:taupoisson}
\end{equation}
In the case of single exponent power law model, see
\cite{ref13,PhysA2007}, when PDF of $n$ is $P(n)\sim
n^{-\lambda}\exp\{-(n_{\mathrm{0}}/n)^m\}$, the distribution
$P(\tau_{k})$ of intertrade time $\tau_{k}$ in $k$-space has the
integral form
\begin{equation}
P(\tau_{k})=C\int_{0}^{\infty}\exp(-n\tau_{k})n^{1-\lambda}
\exp\left[-\left(\frac{n_0}{n}\right)^m\right]\mathrm{d}
n,\label{eq:taupdistrib}
\end{equation}
with $C$ defined from the normalization,
$\int_{0}^{\infty}P(\tau_{k})\mathrm{d} \tau_{k}=1$. The explicit expressions
of the integral (\ref{eq:taupdistrib}) are available for the integer values of
$m$. When $m=1$, PDF (\ref{eq:taupdistrib}) is expressed through the Bessel
function of the second kind whereas for $m>1$ the more complicated structures
of distribution $P(\tau_{k})$ expressed in terms of hypergeometric functions
arise.

Now we have the complete set of equations defining the stochastic
model of the trading activity in the financial markets. We
proposed this model following our increasing interest in the
stochastic fractal point processes
\cite{FPProc,ref16,ref17,ref18}. Our objective to reproduce in
details statistics of trading activity  conditions rather
complicated form of the SDE (\ref{eq:nstoch2}) and low expectation
of analytical results. In this paper we focus on the numerical
analysis and direct comparison of the model with the empirical
data. In order to achieve more general description of statistics
for different stocks we introduce the scaling to
Eq.~(\ref{eq:nstoch2}) with scaled time $t_s=\sigma^2n_0^3t$,
scaled rate $x=n/n_0$ and $\varepsilon'=\varepsilon n_0$.  Then
Eq.~(\ref{eq:nstoch2}) becomes
\begin{equation}
\mathrm{d}x=\left[(1-\gamma_{\sigma})+\frac{m}{2}x^{-m}\right]
\frac{x^4}{(x\varepsilon'
+1)^2}+\frac{x^{5/2}}{(x\varepsilon'+1)}\mathrm{d}W_s.
\label{eq:nscaled}
\end{equation}
The parameter $n_0$ specific for various stocks is now excluded from the SDE
and we have only three parameters to define from the empirical data of trading
activity in the financial markets. We solve Eq.~(\ref{eq:nscaled}) using the
method of discretization.  Introducing the variable step of integration $\Delta
t_s=h_k=\kappa^2/x_k$, the differential equation (\ref{eq:nscaled}) transforms
to the difference equation
\begin{eqnarray}
x_{k+1} &=& x_k+\kappa^2\left[(1-\gamma_{\sigma})+\frac{m}{2}x_k^{-m}\right]
\frac{x_k^3}{(x_k\epsilon'
+1)^2}+\kappa\frac{x_k^{2}}{(x_k\epsilon'+1)}\varepsilon_k ,\\
t_{k+1} & = & t_k+\kappa^2/x_k
\label{eq:difference}
\end{eqnarray}
with $\kappa\ll1$ being a small parameter and $\varepsilon_k$ defining
Gaussian noise with zero mean and unit variance .

With the change of variables $\tau=1/n$ one can transform
Eq.~(\ref{eq:nstoch2}) into
\begin{equation}
\mathrm{d}\tau=\sigma^2\left[\gamma_{\sigma}-\frac{m}{2}\left(\frac{\tau}{
\tau_0}\right)^m\right]\frac{1}{(\epsilon+\tau)^2}dt+\sigma\frac{\sqrt{
\tau}}{\epsilon+\tau}\mathrm{d}W
\label{eq:taustoch}
\end{equation}
with limiting time $\tau_0=1/n_0$.  We will show that this form of driving SDE
is more suitable for the numerical analysis. First of all, the powers of
variables in this equation are lower and the main advantage is that the
Poissonian-like process can be included into the procedure of numerical
solution of SDE. We introduce a scaling of Eq.~(\ref{eq:taustoch}) with the
nondimensional scaled time $t_s=t/\tau_0$, scaled intertrade time
$y=\tau/\tau_0$ and $\epsilon'=\epsilon/\tau_0$. Then Eq.~(\ref{eq:taustoch})
becomes
\begin{equation}
\mathrm{d}y=\frac{\sigma^2}{\tau_0^2}
\left[\gamma_{\sigma}-\frac{m}{2}y^m\right]
\frac{1}{(\epsilon'+y)^2}\mathrm{d}t_s+\frac{\sigma}{\tau_0}
\frac{\sqrt{y}}{\epsilon'+y}\mathrm{d}W_s .
\label{eq:tauscaled}
\end{equation}
As in the real discrete market trading we can choose the
instantaneous intertrade time $y_k$ as a step of numerical
calculations, $h_k=y_k$, or even more precisely as the random
variables with the exponential distribution $P(h_k)=1/y_k
\exp(-h_k/y_k)$. Then we get the iterative equation resembling
tick by tick trades in the financial markets,
\begin{equation}
y_{k+1}=y_k+\frac{\sigma^2}{\tau_0^2}\left[\gamma_{\sigma}-\frac{m}{
2}y_k^m\right]\frac{h_k}{(\epsilon'+y_k)^2}+\frac{\sigma}{\tau_0}
\frac{\sqrt{y_k h_k}}{
\epsilon'+y_k}\varepsilon_k .
\label{eq:tauiterat}
\end{equation}
In this numerical procedure the sequence of $1/y_k$ gives the
modulating rate and the sequence of $h_k$ is the Poissonian-like
intertrade times. Seeking higher precision one can use the
Milshtein approximation instead of Eq.~(\ref{eq:tauiterat}).

\section{Analysis of empirical stock trading data}

We will analyze the tick by tick trades of 26 stocks on NYSE traded for 27
months from January, 2005. An example of the empirical histograms of $\tau_k$
and $N(t,\tau_d)$ and power spectrum of IBM trade sequence are shown on figure
1.  We will adjust the parameters of the Poissonian-like process driven by SDE
Eq.~(\ref{eq:nstoch2}) or Eq.~(\ref{eq:tauiterat}) to reproduce numerically the
empirical trading statistics.

\begin{figure}
\centering
\includegraphics[width=.45\textwidth]{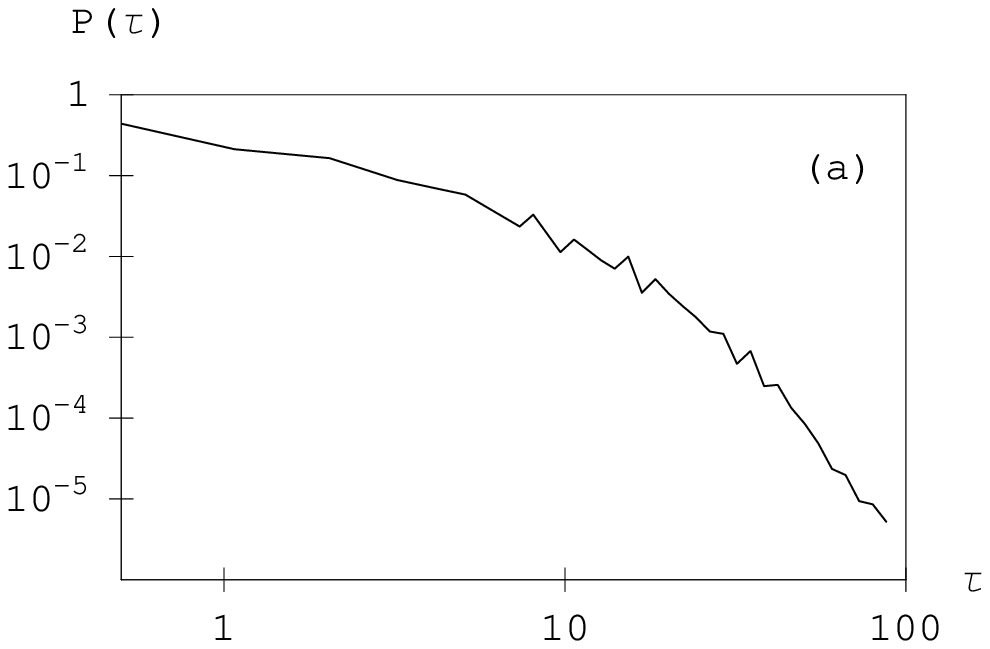}
\includegraphics[width=.45\textwidth]{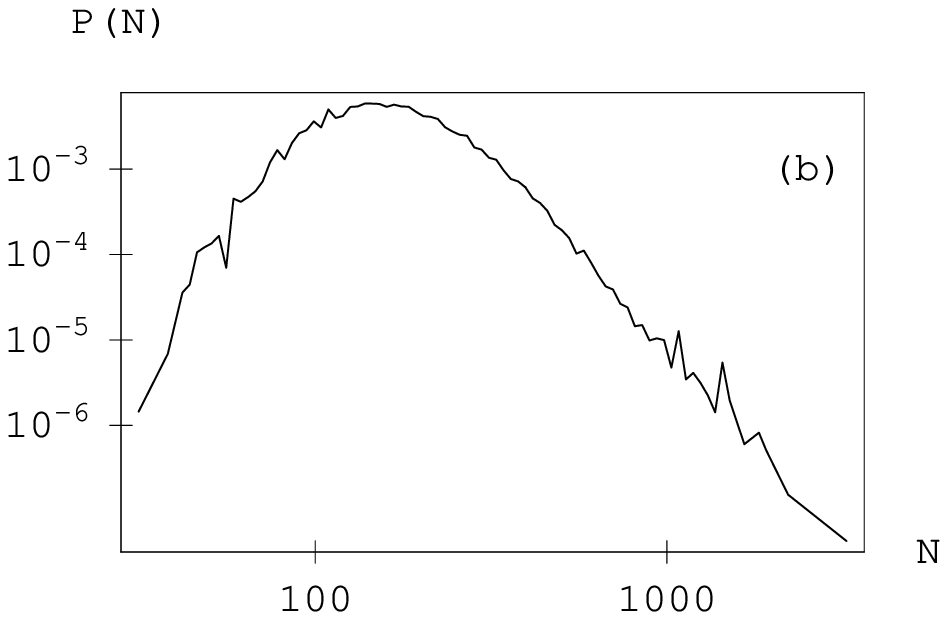}
\includegraphics[width=.45\textwidth]{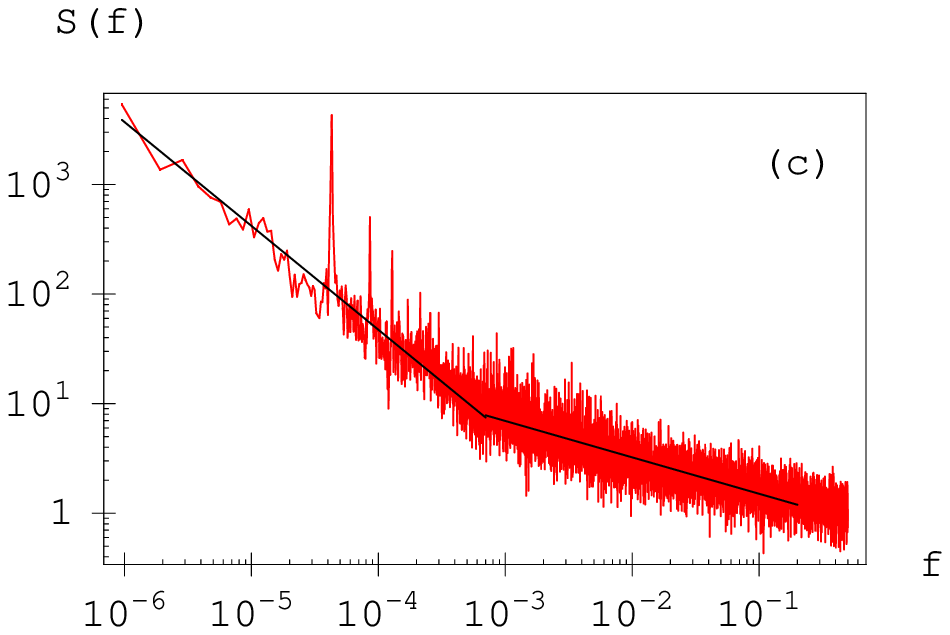}

\caption{Trading statistics of IBM stocks. a) Histogram $P(\tau)$ of the
intertrade time $\tau_k$ sequence; b) Histogram of trading activity $P(N)$
calculated in the time interval $\tau_d=10\,\mathrm{min}$; c) Power spectral
density $S(f)$ of the sequence of trades, straight lines approximate power
spectrum $S(f)\sim1/f^{\beta_{1,2}}$ with $\beta_{1}=0.33$ and $\beta_{2}=0.94
$.}
\label{fig:1}
\end{figure}

The histograms and power spectrum of the sequences of trades for all 26 stocks
are similar to IBM shown on Fig.~\ref{fig:1}. From the histogram $P(\tau_k)$ we
define a model parameter $\tau_0$ for every stock. One can define the exponent
$\lambda'$ from the power-law tail of the histogram $P(N)\sim N^{-\lambda'}$.
The power spectrum exhibits two scaling exponents $\beta_1$ and $\beta_2$ when
approximated by power-law $S(f)\sim1/f^{\beta_{1,2}}$. The empirical values of
$\beta_1$, $\beta_2$, $\tau_0$ and $\lambda'$ for $\tau_d=10\,\mathrm{min}$ are
presented in Table \ref{tab1}.

\begin{table}
\caption{The empirical values of $\tau_0$, $\beta_1$, $\beta_2$ and $\lambda'$.}
\label{tab1}

\begin{tabular}{|c|c|c|c|c||c|c|c|c|c|}
\hline
Stock & $\tau_{0}$ & $\beta_{1}$ & $\beta_{2}$ & $\lambda'$ & Stock & $\tau_{0}$ & $\beta_{1}$ & $\beta_{2}$ & $\lambda'$ \\
\hline
ABT & 7  & 0.27  &  0.92  &  5.8 & JNJ & 4 & 0.31 & 0.87 & 4.4\\
ADM & 8 &  0.29 & 0.96 &  3.5 & JPM & 4.5 & 0.3 & 0.8 & 5.7\\
BA & 8 & 0.32 & 0.84 & 4.5 & KO & 6 & 0.29 & 0.85 & 6.5\\
BMY  & 7  & 0.3 & 0.9 & 4.4 & LLY & 8 & 0.32 & 0.84 & 4.5\\
C &  3 & 0.31 & 0.87  & 4.6 & MMM & 8 & 0.26 & 0.96 & 4.8\\
CVX  & 4  & 0.3 & 0.87 & 6.4 & MO & 5 & 0.36 & 0.9 & 3.6\\
DOW & 7 &  0.32 & 0.88 & 5.7 & MOT & 4 & 0.29 & 0.93 & 3.4\\
FNM  & 9  & 0.36 & 0.95 & 3.4 & MRK & 5 & 0.35 & 0.82 & 3\\
FON  & 8 &  0.29 & 0.97 & 3.4 & SLE & 10 & 0.2 & 0.75 & 4.5\\
GE & 2.25 & 0.27 & 0.87 & 4.7 & PFE & 2 & 0.31 & 0.96 & 3.6\\
GM & 6 & 0.36 & 0.93 &2.7 & T & 10 & 0.28 & 0.88 & 3.7\\
HD & 4 & 0.32 & 0.95 & 5.4 & WMT & 4 & 0.31 & 0.84 & 4.7\\
IBM & 5 & 0.3 & 0.87 & 4.1 & XOM & 3 & 0.34 & 0.9 & 4.4\\
\hline
\multicolumn{6}{|l|}{\textbf{Average}} &
\textbf{5.8} & \textbf{0.305} & \textbf{0.89} & \textbf{4.4}\\
\hline
\end{tabular}

\end{table}

\begin{figure}
\centering
\includegraphics[width=.45\textwidth]{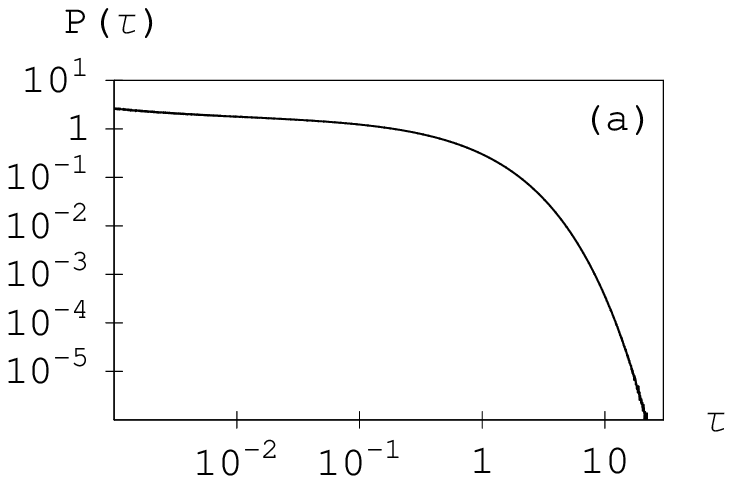}
\includegraphics[width=.45\textwidth]{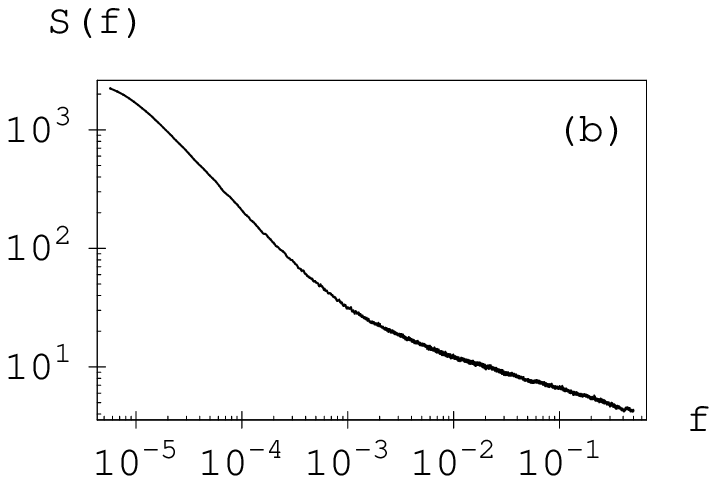}

\caption{Distribution of the Poissonian-like interevent time $\tau_k=h_k$, (a),
and power spectrum, (b), of the sequence of point events calculated from
Eq.~(\ref{eq:tauiterat}) with the adjusted parameters $\sigma'=0.006$,
$\gamma_{\sigma}=0.64$, $\epsilon'=0.05$.}
\label{fig:2}
\end{figure}

Values of $\beta_1$ and $\beta_2$ fluctuate around $0.3$ and $0.9$,
respectively, as in the separate stochastic model realizations. The crossover
frequency $f_c$ of two power-laws exhibits some fluctuations around the value
$f_c\approx10^{-3}\,\mathrm{Hz}$ as well.  One can observe considerable
fluctuations of the exponent $\lambda'$ around the mean value $4.4$. Notice
that value of exponent $\lambda'$ for integrated trading activity $N$ is higher
than for $n$. Our analysis shows that the explicit form of the $P(\tau_{k})$,
Eq.~(\ref{eq:taupdistrib}) with $m=2$ and $\lambda=2.7$, fits empirical
histogram of $\tau_k$ for all stocks very well and fitting parameter $\tau_0$
can be defined for every stock. Values of $\tau_0$ are presented on Table
\ref{tab1}.  From the point of view of the proposed model the parameter
$\tau_0$ is specific for every stock and reflects the average trading intensity
in the calm periods of stock exchange. We eliminate these specific differences
in our model by scaling transform of Eq.~(\ref{eq:taustoch}) arriving to the
nondimensional SDE (\ref{eq:tauscaled}) and its iterative form
(\ref{eq:tauiterat}). These equations and parameters $\sigma'=\sigma /\tau_0$,
$\gamma_{\sigma}$, $\epsilon'$ and $m=2$ define our model, which has to
reproduce in details power-law statistics of the trading activity in the
financial markets. From the analysis based on the research of fractal
stochastic point processes \cite{ref13,PhysA2007,FPProc,ref16,ref17,ref18} and
by fitting the numerical calculations to the empirical data we arrive at the
collection of parameters $\sigma'=0.006$, $\gamma_{\sigma}=0.64$,
$\epsilon'=0.05$. In figure \ref{fig:2} we present histogram of the sequence of
$\tau_k=h_k$, (a), and the power spectrum of the sequence of trades as point
events, (b), generated numerically by Eq.~(\ref{eq:tauiterat}) with the
adjusted parameters.

For every selected stock one can easily scale the model sequence of intertrade
times $\tau_k=h_k$ by empirically defined $\tau_0$ to get the model sequence of
trades for this stock. One can scale the model power spectrum $S(f)$ by
$1/\tau_0^2$ for getting the model power spectrum $S_{\mathrm{stock}}(f)$ for
the selected stock $S_{\mathrm{stock}}(f)=S(f)/\tau_0^2$.
\begin{figure}
\centering
\includegraphics[width=.45\textwidth]{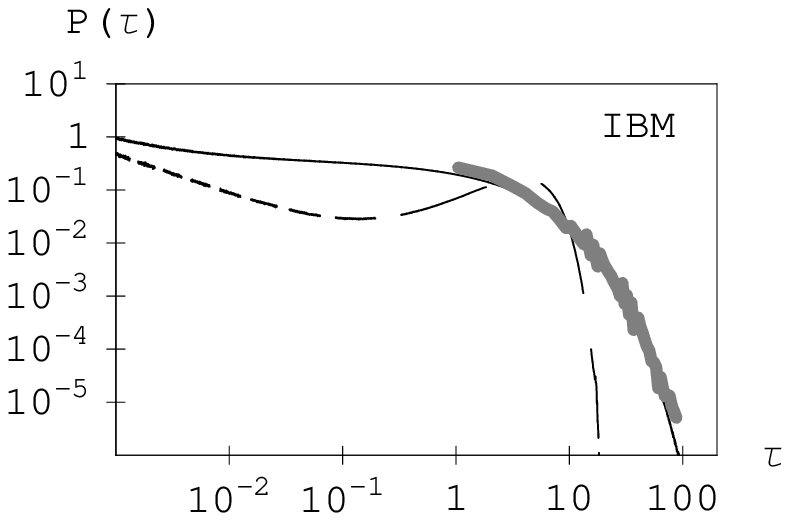}
\includegraphics[width=.45\textwidth]{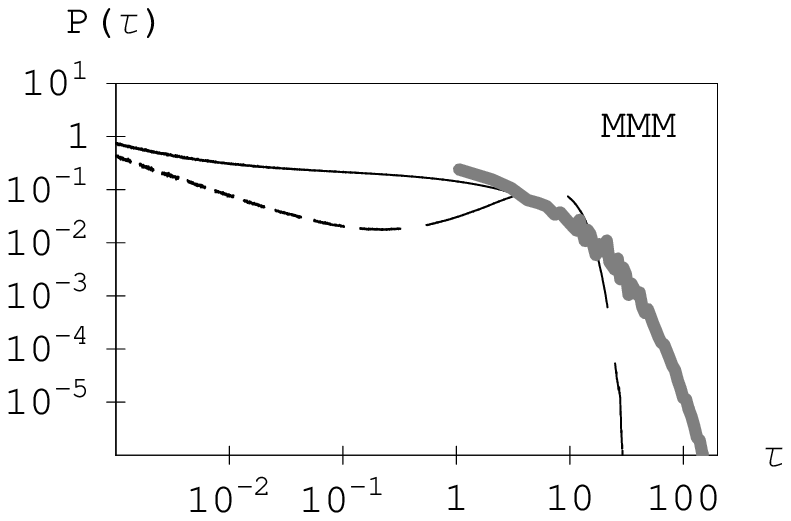}

\caption{Distribution $P(\tau)$ of intertrade time $\tau$ for IBM
and MMM stocks; empirical histogram, tick gray line, modeled
Poissonian-like distribution, solid line, distribution of driving
$\tau=y_k$ in Eq. (\ref{eq:tauiterat}), dashed line. } \label{fig:3}
\end{figure}

\begin{figure}
\centering
\includegraphics[width=.45\textwidth]{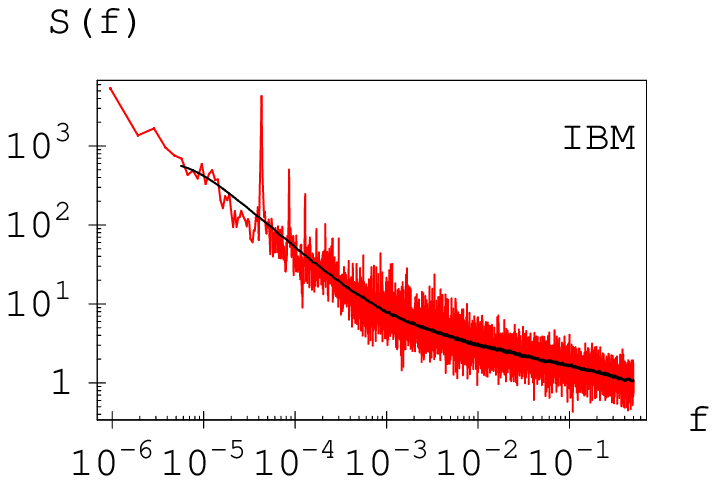}
\includegraphics[width=.45\textwidth]{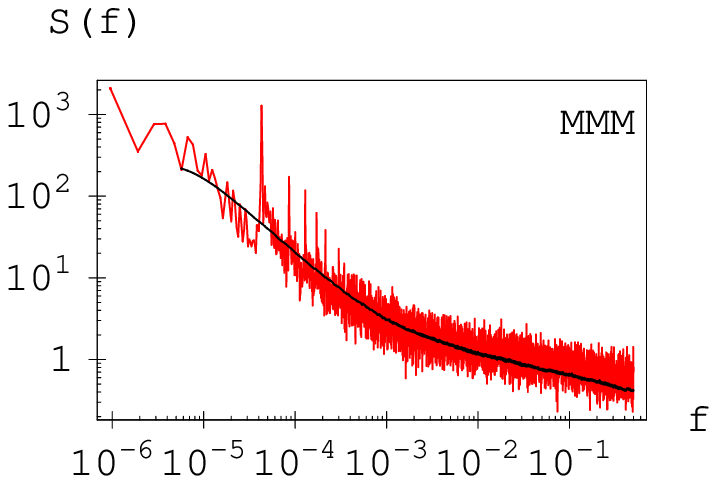}

\caption{Modeled, smooth curves, and empirical, sharp curves, power spectra of
trading activity for IBM and MMM stocks.}
\label{fig:4}
\end{figure}

\begin{figure}
\centering
\includegraphics[width=.45\textwidth]{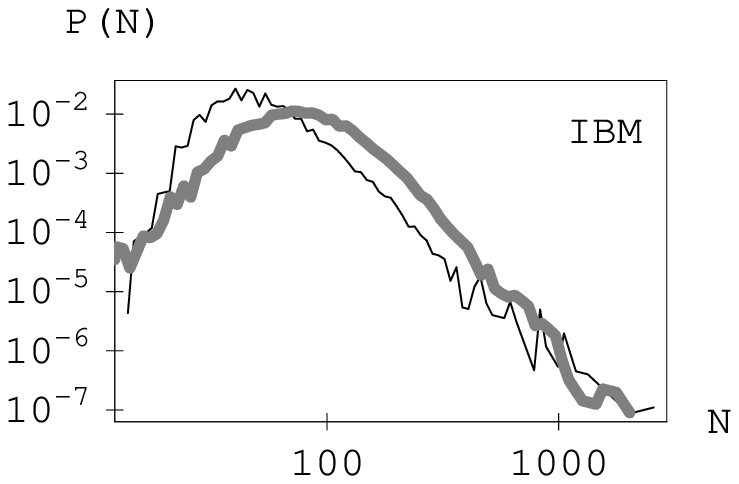}
\includegraphics[width=.45\textwidth]{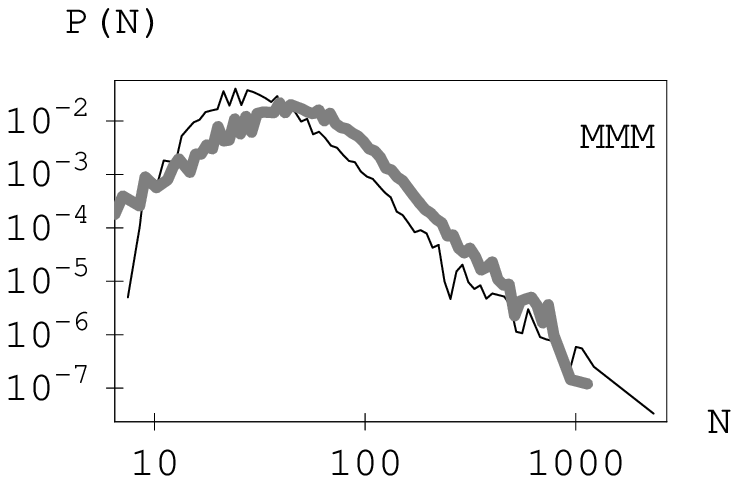}

\caption{Modeled, thin line, and empirical, tick line, trading
activities for IBM and MMM stocks in the time interval
$\tau_{\mathrm{d}}=300s$.} \label{fig:5}
\end{figure}

We proposed the iterative Eq.~(\ref{eq:tauiterat}) as quite accurate stochastic
model of trading activity in the financial markets. Nevertheless, one has to
admit that real trading activity often has considerable trend as number of
shares traded and the whole activity of the markets increases. This might have
considerable influence on the empirical long range distributions and power
spectrum of the stocks in consideration. The trend has to be eliminated from
the empirical data for the detailed comparison with the model. Only few stocks
from the selected list in Table \ref{tab1} have stable trading activity in the
considered period. In figures \ref{fig:3}, \ref{fig:4} and \ref{fig:5} we
provide the comparison of the model with the empirical data of those stocks. As
we illustrate in figure \ref{fig:3}, the model Poissonian-like distribution can
be easily adjusted to the empirical histogram of intertrade time, with
$\tau_0=5\,\mathrm{s}$ for IBM trade sequence and with $\tau_0=8\,\mathrm{s}$
for MMM trading. The comparison with the empirical data is limited by the
available accuracy, $1\,\mathrm{s}$, of stock trading time $t_k$. The
probability distribution of driving $\tau=y_k$ Eq.~(\ref{eq:tauiterat}), dashed
line, illustrates different market behavior in the periods of the low and high
trading activity. The Poissonian nature of the stochastic point process hides
these differences by considerable smoothing of the PDF.  Figure \ref{fig:4}
illustrates that the long range memory properties of the trading activity
reflected in the power spectrum are universal and arise from the scaled driving
SDE (\ref{eq:nscaled}) and (\ref{eq:tauscaled}). One can get power spectrum of
the selected stock trade sequence scaling model spectrum, figure \ref{fig:2}
(b), with $1/\tau_0^2$. The PDF of integrated trading activity $N$ is more
sensitive to the market fluctuations.  Nevertheless, as we demonstrate in
figure \ref{fig:5}, the model is able to reproduce the power-law tails very
well.

\section{\label{sec:concl}Conclusions}

We proposed the generalization of the point process model as the
Poissonian-like sequence with slowly diffusing mean interevent time
\cite{PhysA2007} and adjusted the parameters of the model to the empirical data
of trading activity in the financial markets. A new form of scaled equations
provides the universal description with the same parameters applicable for all
stocks. The proposed new form of the continuous stochastic differential
equation enabled us to reproduce the main statistical properties of the trading
activity and waiting time, observable in the financial markets. In proposed
model the fractured power-law distribution of spectral density with two
different exponents arise. This is in agreement with the empirical power
spectrum of the trading activity and volatility and implies that the market
behavior may be dependent on the level of activity. One can observe at least
two stages in market behavior: calm and excited. Ability to reproduce empirical
PDF of intertrade time and trading activity as well as the power spectrum in
every detail for various stocks provides a background for further stochastic
modeling of volatility.

\noindent \textbf{Acknowledgment}

We acknowledge the support by the Agency for International Science and
Technology Development Programs in Lithuania and EU COST Action P10 ``Physics
of Risk''.

\end{document}